\documentclass[iop]{emulateapj}

\newcommand{\mbh}{M$_{\rm BH}$}   
\newcommand{\Hb}{H$\beta$}
\newcommand{\Ha}{H$\alpha$}

\newcommand{\ergs}{erg s$^{-1}$}

\begin{document}

\title{Seoul National University AGN Monitoring Project.  I. Strategy and Sample}


\author{Jong-Hak Woo$^{1}$}
\author{Donghoon Son$^{1}$}
\author{Elena Gallo$^{2}$}
\author{Edmund Hodges-Kluck$^{2,3}$}
\author{Jaejin Shin$^{1}$}
\author{Yiseul Jeon$^{1}$}
\author{Hyun-Jin Bae$^{1}$}
\author{Hojin Cho$^{1}$}
\author{Wanjin Cho$^{1}$}
\author{Daeun Kang$^{1}$}
\author{Wonseok Kang$^{6}$}
\author{Marios Karouzos$^{1}$}
\author{Minjin Kim$^{4}$}
\author{Taewoo Kim$^{6}$}
\author{Huynh Anh N. Le$^{1}$}
\author{Daeseong Park$^{5}$}
\author{Songyoun Park$^{1}$}
\author{Suvendu Rakshit$^{1}$}
\author{Hyun-il Sung$^{5}$}


\affil{$^{1}$Astronomy Program, Department of Physics and Astronomy, Seoul National University, Seoul, 08826, Republic of Korea\\
$^{2}$Department of Astronomy, University of Michigan, Ann Arbor, MI 48109, USA\\
$^{3}$NASA/GSFC, Code 662, Greenbelt, MD 20771, USA\\
$^{4}$Department of Astronomy and Atmospheric Sciences, Kyungpook National University, Daegu 41566 Korea\\
$^{5}$Korea Astronomy and Space science Institute, Daejeon, Republic of Korea\\
$^{6}$National Youth Space Center, Goheung, Jeollanam-do, 59567, Korea
}


\begin{abstract}
While the reverberation mapping technique is the best available method for measuring black hole mass 
in active galactic nuclei (AGN) beyond the local volume, this method has been mainly applied to relatively low-to-moderate luminosity AGNs
at low redshift. We present the strategy of the Seoul National University AGN Monitoring Project, which aims at 
measuring the time delay of the \Hb\ line emission with respect to AGN continuum, using a sample of relatively high luminosity AGNs out to z $\sim$0.5. 
We present simulated cross correlation results based on a number of mock light curves, in order to optimally determine monitoring duration and cadence. We describe our campaign strategy based on the simulation results and the availability of observing facilities. 
We present the sample selection, and the properties of the selected 100 AGNs, including the optical luminosity, expected time lag, black hole mass, and Eddington ratio. 
\end{abstract}


\section{INTRODUCTION} \label{section:intro}

Black hole mass (\mbh) is a key parameter in understanding the physics of active galactic nuclei (AGN). 
The correlation of \mbh\ with the global properties of host galaxies observed in inactive
\citep[e.g.,][]{FerrareseMerritt2000, Gebhardt2000, KormendyHo2013} and active galaxies \citep[e.g.,][]{Woo+13, Grier+13, Woo+15} are often interpreted as a strong connection between black hole growth and galaxy evolution \citep[see][]{KormendyHo2013}. In order to investigate the physics of AGN phenomena \citep[e.g.,][]{Woo+02, Kollmeier2006,Davis2007,Bentz+13,Woo+16}, \mbh\ must be accurately determined. Dynamical methods based on high angular resolution data are a common approach to measure \mbh\ (e.g., Kormendy \& Ho 2013). However, owing to the small scale of the sphere of the influence of typical supermassive black holes, the dynamical method is limited to galaxies within a distance of $\sim$100 Mpc with the current technology. 

Reverberation mapping is the best available method to determine \mbh\ beyond the local volume.  Via the virial assumption that the kinematics of the gas in the broad-line region (BLR) is governed by black hole's gravity, \mbh\ can be determined by measuring the size and velocity of the BLR gas. The geometry and kinematics of the BLR gas can be mapped in time domain with the reverberation mapping technique \citep{Blandford1982,Peterson1993}, as \mbh\ $\propto$ R$_{BLR}$ V$^2$, where R$_{BLR}$ is the measured time lag (i.e., light travel time) between AGN continuum region and BLR gas, while V is velocity measured from the width of the broad-emission lines. 
 Previous reverberation studies showed that the size (time lag) and velocity of the \Hb\ emission line measured at different epochs clearly followed the virial relation, indicating that the virial assumption is valid \citep{Peterson2004}, 
while the black hole mass suffers systematic uncertainty due to the complex nature of the BLR kinematics \citep{Park+12, Bentz+10, Pancoast+14}.

More than 60 masses have been reliably measured using this method \citep[e.g.,][]{Wandel1999, Kaspi+00,Peterson2004,Bentz+09,Barth2011,Grier+13, Barth2015,Fausnaugh+17a, Park+17}. However, due to observational challenges, this method has been mainly applied to relatively low-to-moderate luminosity AGNs at low redshift. As a long-term spectroscopic monitoring campaign is required for reliable lag measurements, it is difficult to increase the sample size. 

Nevertheless, recent reverberation campaigns have achieved significant progress by increasing the size and dynamic range of the sample. For example, the Lick AGN Monitoring Project 2008, 2011 utilized the 3-m aperture telescope to carry out semi-consecutive nightly observations over 2-3 months, respectively, in 2008 and 2011. This program was the first dedicated reverberation mapping project using a mid-size telescope \citep{Barth2015}. 
Other studies targeted high Eddington ratio AGNs for investigating the photoionization and validity of the indirect mass estimators \citep{Du2015, Du2016}. 
Based on a very long-term campaign, a number of lag measurements using the rest-frame UV lines has been also reported for high-z AGNs \citep{Lira+18}. Recently, a new approach has been applied with a multi-object spectrograph, for simultaneously monitoring multiple AGNs. For example, the SDSS-reverberation mapping project has been carried out over 6 months by monitoring hundreds of AGNs out to z$\sim$4.5 \citep{Shen+16, Grier+17}.  Also, as a part of the Australian spectroscopic dark energy survey (OzDES), a multi-object reverberation program is underway for monitoring hundreds of quasars.

Reverberation mapping data are of paramount importance since they provide the fundamental calibration for all indirect \mbh\ estimators, which in turn enable us to investigate black hole growth and accretion physics over cosmic time. However, the reverberation mapping method is still limited to relatively small number of local AGNs, while numerous AGNs are detected out to z$\sim$7. 
The main limitation of current reverberation studies is the lack of high-luminosity and high-redshift AGNs in the sample. The majority of the \Hb\ reverberation-mapped sample are low to intermediate luminosity AGNs, predominantly located at z $<$$\sim$0.1. 
The bias toward low luminosity AGNs is a natural consequence of the lack of long-term monitoring programs since several years of spectroscopic observations are required to measure large time-lags (i.e, $>$ 100 light days in the rest-frame). 
In addition, time dilation due to Hubble expansion increases the observed lag of high-z  AGNs by a factor of (1+z). Thus, a very long monitoring is required for reverberation mapping of high-z AGNs \citep[e.g.,][]{Lira+18}. So far, the longest observed \Hb\ lag is $\sim$300 light days in the rest-frame. 

The luminosity limitation of the current reverberation-mapped AGN sample can seriously bias the empirical size-luminosity relation \citep{Bentz+06}, which is mainly used to estimate \mbh\ of a large number of AGNs. Thus, a major effort to obtain relatively long lag measurements using luminous AGNs is necessary to overcome the current limitation of \mbh\ studies.  In this paper, we present the Seoul National University AGN monitoring project (SAMP), which aims at measuring \Hb\ time lag of relatively high luminosity AGNs at z $<$ 0.5. The strategy of the project and the simulation results are presented in Section 2 and Section 3, respectively. The sample selection and sample properties are presented in Section 4. Summary follows in Section 5. 

\section{Project strategy}

High-luminosity AGNs will be the most important targets to overcome the current limitation of the \Hb\ reverberation studies.
Since the expected time lag of high luminosity AGNs is very long, not to mention the time dilation effect due to the redshift, several years of monitoring is required as a minimum. At the same time, a relatively large aperture telescope is needed to measure the flux variation of the \Hb\ emission line of these faint AGNs at cosmological distances. 

To set up a long-term monitoring project, we consider three factors: expected lags, monitoring duration and cadence, which are the key parameters for a successful campaign. The expected lag depends on the luminosity of each target and can be estimated from the size-luminosity relation \citep{Bentz+06}. 
We select a sample of relatively high luminosity AGNs with a monochromatic luminosity at 5100\AA\ L$_{5100}$$\sim$10$^{44-46}$ \ergs, in order to cover the high-luminosity end of the size-luminosity relation. The \Hb\ lag of these AGNs  is expected to be more than $\sim$100 light days in the observed-frame. Based on a three year monitoring program, we expect to measure \Hb\ time lags up to $\sim$300 light days, by considering that the monitoring length has to be longer than the expected lag by a factor of 3-5. Thus, we aim at measuring \Hb\ lag of 100-400 light days in the observed-frame. 
In terms of the cadence, it is desirable to have a factor of 5 or better time resolution for a given time lag, as performed in the previous studies. 
Thus, a $\sim$20 day cadence is required to measure a 100 light day time lag. 

Considering these conditions, we initially design a three year monitoring program, which is also limited by the funding and availability of observing facilities. For spectroscopy, we need $\sim$20 nights per year of a mid-size telescope, in order to obtain good quality spectra. With a 20-30 day cadence, a total of $\sim$50-60 epochs of spectroscopic data can be obtained during a three year campaign. In the case of photometry, we can achieve a higher time resolution, i.e., a 10 day cadence since photometry can be well performed with smaller telescopes, which are much easier to access.

\section{simulation results}

\begin{figure}
\centering
\includegraphics[width = 0.49\textwidth]{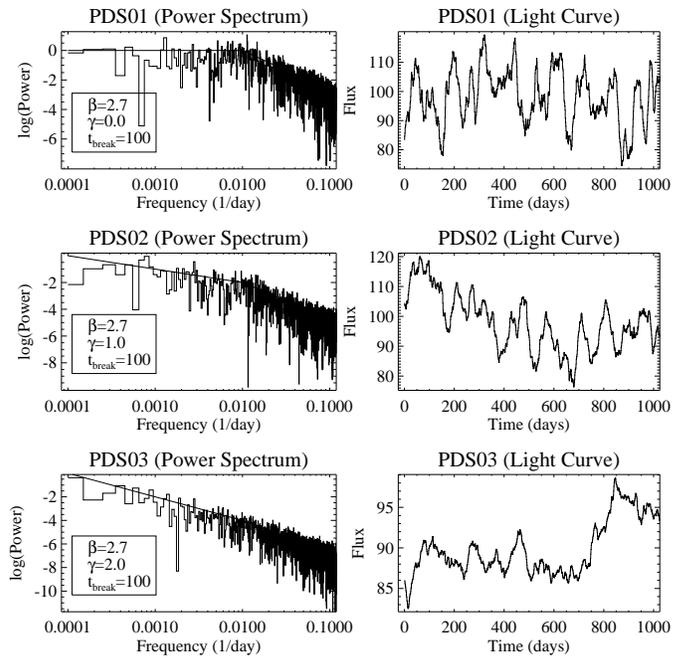}
\caption{Examples of power density spectra generated with a broken power law (left)
and the corresponding light curves (right). 
The three parameters, $\beta$, $\gamma$, and t$_{\rm break}$ are specified in the inner panel. 
The light curves are normalized to an rms amplitude of 10\%.
}
\end{figure}

To quantify the best strategy for the monitoring program, we simulate the success rate as a function of time lag, campaign duration, and cadence. We generate continuum and line emission light curves to perform cross-correlation analysis (Section 3.2).
While it is clear that a shorter cadence and a longer campaign duration can increase the success rate,  
we need to investigate optimal criteria of cadence and duration since observing facilities are limited.  
By comparing the input lag, which is used to delay an emission line light curve with respect to a continuum light curve,
with the measured lag based on the cross-correlation analysis, we determine whether the lag measurement is successful.  
The simulation is conducted with various cadences and campaign durations to calculate the success rate for each case (Section 3.3).
After we finalize the optimal criteria, which satisfy success rate $\ge$ 80\%, we select the sample accordingly (Section \ref{sel}).

\subsection{Generating light curves}\label{sim_lc}

\begin{figure*}
\centering
\includegraphics[width = 0.85\textwidth]{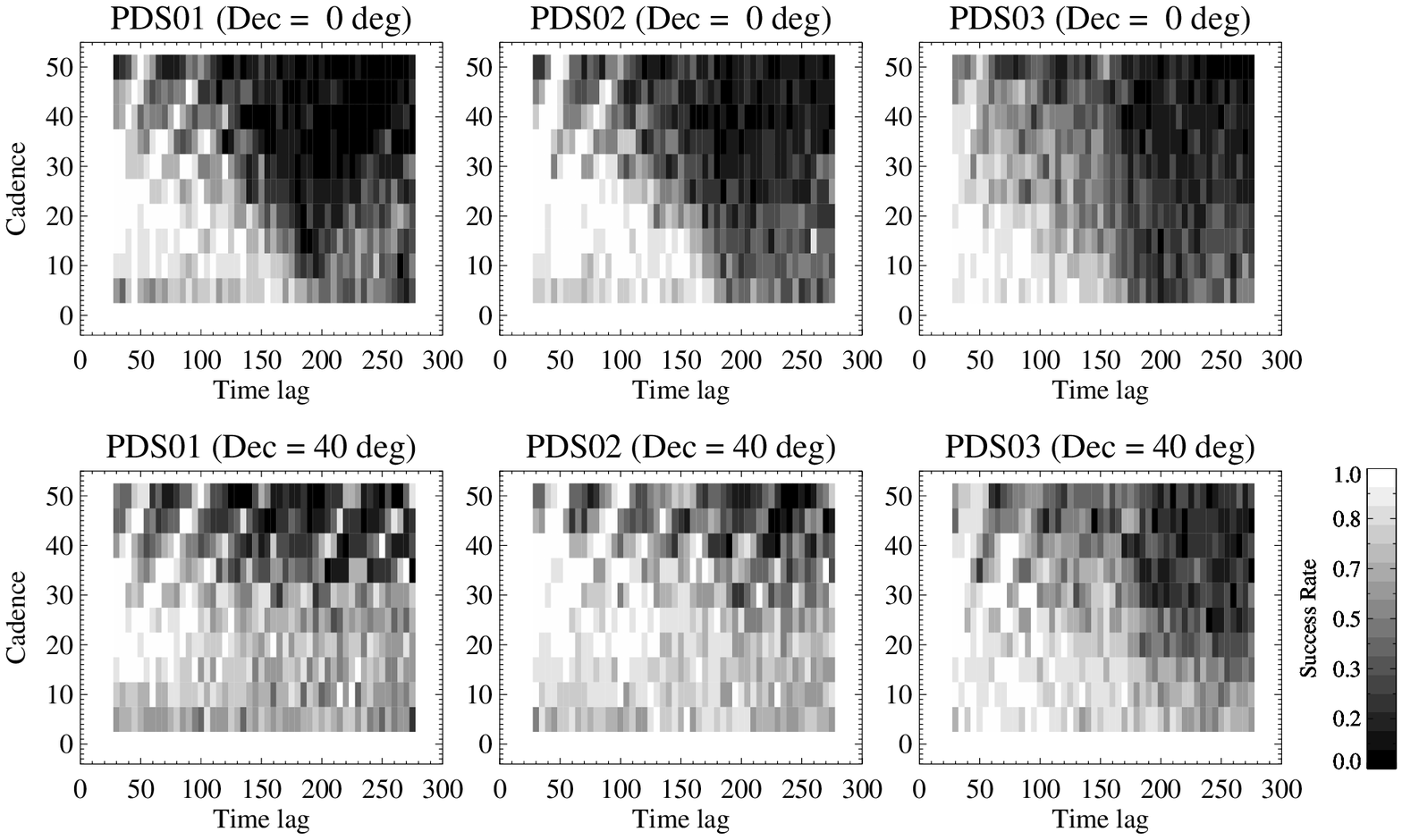}
\caption{
The estimated success rate maps of three PDSs as a function of cadence and time lag for the three year campaign duration.
We assumed two different seasonal gaps (i.e., Dec = 0 and 40 degree).
The colors indicate success rate from 0\% (black) to 100\% (white). }
\end{figure*}

\begin{figure*}
\centering
\includegraphics[width = 0.85\textwidth]{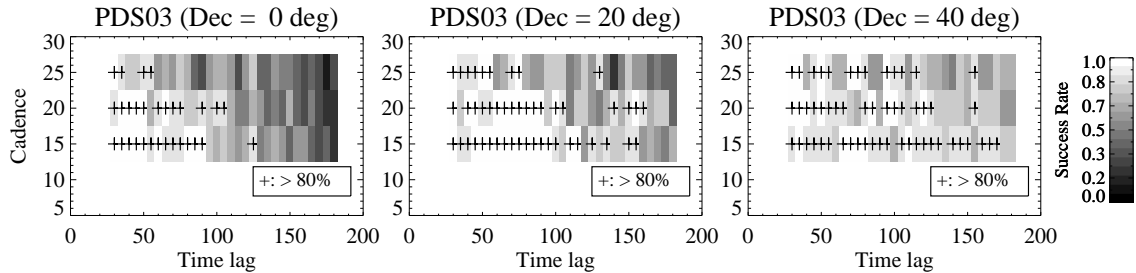}
\caption{
Simulated success rate maps for three declination values (i.e., Dec = 0, 20, and 40 deg) for the case of PDS03 and three year campaign duration.
The plus sign indicates a set of a cadence and time lag, that satisfies the success rate greater than 80\%. 
The colors indicate success rate from 0\% (black) to 100\% (white). }
\end{figure*}

We simulate a continuum light curve by generating mock light curves with a various length from
one to five years, using a power density spectrum (PDS).
Then, we delay the continuum light curve by a specific time lag, in order to define an emission-line light curve. 
These two light curves are sampled using a sampling size according to a specific cadence (e.g., 60 epochs for three years).

For generating mock light curves, we use the method described by \citet{Timmer+95}. 
The variability of AGN can be described as a 1/f fluctuation \citep{Lawrence+88}, 
where the 1/f describes the distribution of power as a function of frequency (f) in the power spectrum. 
From Kepler data (Barth, A. private communication), the slopes of PDS for the high frequency region are -2.5 -- -3.0.
However, if we apply this steep slope to the low frequency, the generated light curves show
very strong secular trends, which do not reflect the actual AGN light curves. 
To create relatively flat light curves on the long time scales (i.e., low frequency), 
we use a form of a broken power law, $P(f) \propto f^ {- \beta}$ for f$ >$ 1 / t$_{\rm break}$  
or $P(f) \propto f^{-\gamma}$ for f $<$ 1 / t$_{\rm break}$, 
where frequency ($f$) has a unit of day$^{-1}$, 
$\beta$ is the power law index for the high frequency region, and 
$\gamma$ is the power law index for the low frequency region.
We fix the $\beta$ value to 2.7 which is a middle value between -2.5 and -3.0 from the Kepler data. 
We test various $\gamma$ values to define long term shapes of light curves 
and select three values, i.e., $\gamma=0, 1, 2$. These $\gamma$ values 
represent light curves with a strong trend ($\gamma=0$) 
and a moderate trend ($\gamma=2$). 
We fix the t$_{break}$ to 100 days because 
this value makes the light curves with various $\gamma$ realistic. 
We call these three PDSs as PDS01, PDS02, and PDS03
when $\gamma$=0, 1, and 2, respectively as presented in Figure 1 (left).

Using a set of these parameters, we generate a power spectrum 
with a sampling interval as one day and added randomized Fourier amplitudes to the spectrum. 
Then, via Fourier transformation, we create a series of light curves as shown in Figure 1 (right).
Note that as $\gamma$ value increases, the power of the low frequency (i.e., long-term variation of light curves) gets stronger and the secular trends get weaker. 

For the continuum light curves, we normalize the variability amplitude by adopting a typical value. 
AGN variability studies show that the rms amplitude in the optical continuum is $\sim$10-20\%. 
In this simulation we adopt conservative 10\% variability amplitude.
In the case of emission-line light curves, we assume that the amplitude of variability is the same as in the continuum  
to simplify the simulations.  
After sampling the continuum and the emission line light curves with a specific cadence, 
we assign 1\% errors on the continuum flux and 2\% error on the emission line flux measurements. 

For each target, there will be a seasonal gap due to the lack of observability. The length of the seasonal gap is a function 
of declination of each target for a given ground-base telescope.
For example, at the Lick observatory located at latitude = +37:20:36,
the maximum observable nights with airmass $<$ 2 is 180, 212, 238, and 266 days per year, respectively for a target with Dec = 0, 20, 40, and 60 degree. 
Thus, if we select targets at Dec $>$ 40 degree, the seasonal gap will be less than 4 months. 
 
\subsection{Time lag measurement}\label{sim_jav}

In order to measure time lag, we generate a series of two light curves (continuum and emission line) using a set of four parameters. 
First. we use three different PDSs as presented in Figure 2. Second, we choose an intrinsic time lag between 30 and 275 days. Third,
we fix the monitoring cadence between 5 and 50 days. Four, we determine a seasonal gap based on three different declinations (i.e., Dec = 0, 20, 40 degree). 
To measure the time lag of an emission line light curve with respect to a continuum light curve, 
we utilize JAVELIN, which is an analysis software used for measuring time lag between two light curves by adopting a damped random walk to model AGN variability \citep{Zu+11, Zu+13}.
For each set of parameters, we use 10 random light curves to measure time lag, and calculate the success rate. 
To decide whether  the measured time lag agrees with the intrinsic time lag, we consider the measurement as a success 
if the relative difference between the measured and intrinsic lags is less than 10\%.

\subsection{Success rate} 

We present the success rate for a 3 year campaign as an example, depending on cadence and time lag, in Figure 2. Considering the effect of seasonal gap, we show the results for the case of Dec = 0 and 40 degree, respectively. 
As expected, the light curves generated from a conservative power density spectrum PDS03 (i.e., $\gamma$=2.0, highest power at low frequency) 
generally show lower success rate than those from other PDSs with smaller $\gamma$ values. 
Also, when the seasonal gap is shorter (i.e., Dec = 40 deg), the success rate is higher.
Note that if the intrinsic time lag coincides with the length of the seasonal gap (e.g. a target with a $\sim$180 day lag at Dec = 0 deg), 
the success rate is almost 0\% regardless of the cadence (see top left panel in Figure 2).

In Figure 3, we mark the set of the intrinsic time lag and the cadence when the success rate is larger than 80\%,
by focusing on the case of PDS03 (the most conservative case).
If we adopt a 15 day cadence for a 3 year monitoring program, intrinsic time lags of 30-150 days can be
measured with a 80\% success rate for AGNs at Dec $>$ 20 degree. 
If the cadence is 20 days, we can still measure intrinsic time lags between 30 and 130 days for AGNs at Dec $>$ 40 degree.

To investigate how the campaign duration affects the success rate, we present 
the success rate as a function of cadence and campaign duration, using the light curves from PDS03,
and the fixed intrinsic time lag = 100 days and Dec = 40 degree in Figure 4. 
For this simulation, we generate 50 sets of light curves for each bin (cadence and campaign duration), and 
measure the time lag of each case. The number in each box indicates the success rate in per cent.  
For example, in the case of a 20 day cadence, we obtain the 76\% success rate for a 3 year campaign,
while the success rate increases to 96\% if the campaign duration is extended to 5 years. 

\begin{figure}
\includegraphics[width = 0.36\textwidth]{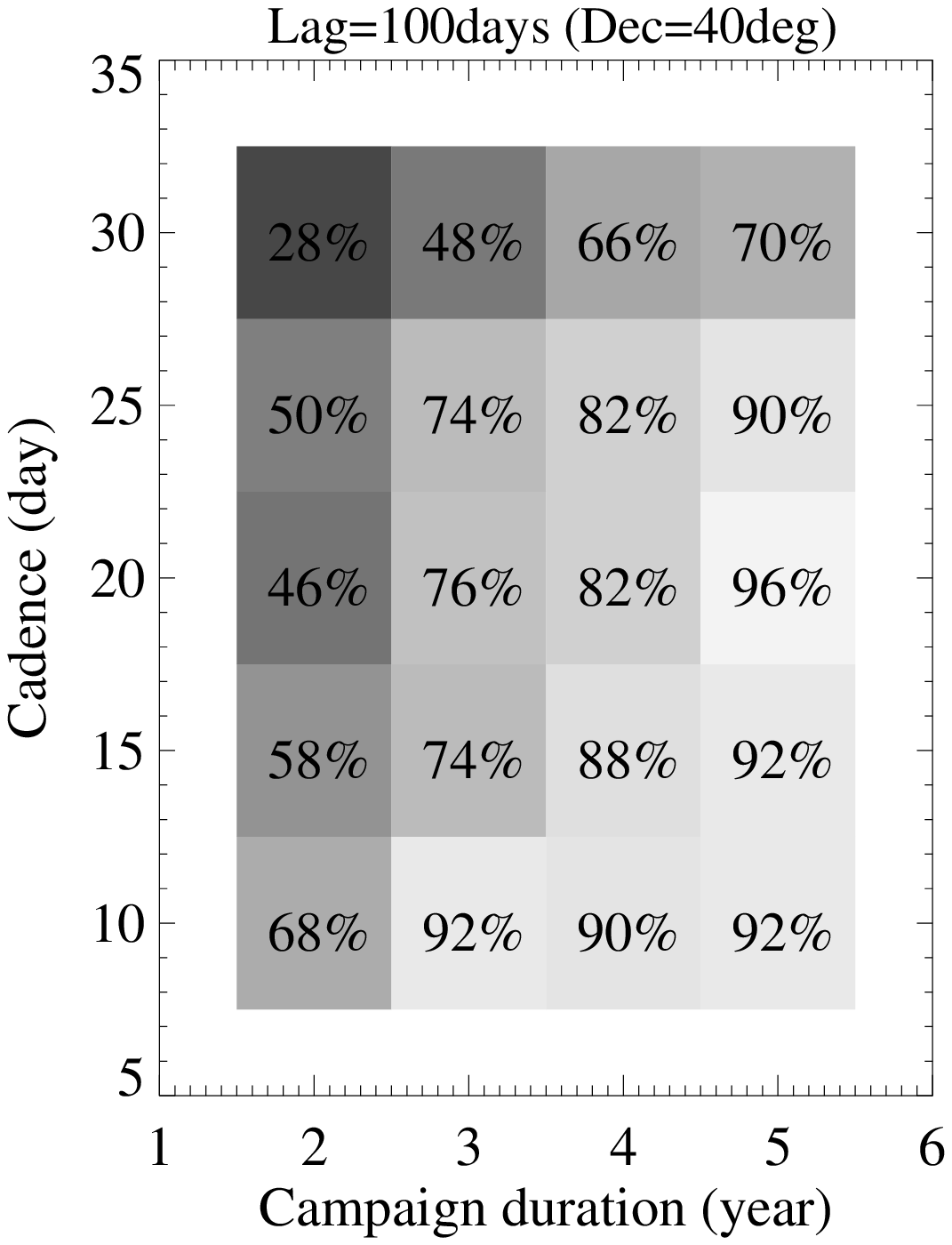}
\caption{
Success rate map as a function of cadence and campaign duration for a specific target with the expected time lag = 100 days and located at Dec = 40 deg. 
The number in each box indicates the success rate in \%. }
\end{figure}

Based on these results, we conclude that a 3 year monitoring campaign with a 20 day cadence 
can provide high quality lag measurements within 10\% error, with a 80\% success rate for AGNs with intrinsic time lag $\sim$150 days. 
Note that the simulated results are based on idealized light curves and there could be various other effects that could decrease the 
success rate. Here we take these results as a guide line for preparing a reasonable monitoring campaign. We decide to perform 
a 3 year monitoring campaign with a $\sim$20 day cadence for spectroscopy and a factor of 2 better time resolution for photometry.

\begin{table*}
\begin{center}
\caption{Properties of our samples\label{tbl_target}}
\begin{tabular}{lllccrcrrr}
\tableline\tableline
  Name & RA & Dec & z & B & (1+z)lag & 	log L$_{5100,spec}$	& (1+z)lag$_{spec}$  & logM$_{\rm BH}$ & Ref.Spec. \\
      &        &          &     &       & (day) & (erg s$^{-1}$)	& (day) & (M$_{\odot}$) &  \\
 (1)  & (2)    & (3)      & (4) & (5)   & (6)   & (7)   			& (8)  	& (9) & (10)	\\
\tableline
Pr1\_ID01		&		01:01:31.1		&		+42:29:36.0		&		0.190		&		16.9		&	84.7	& 44.6	&	87.8	&	8.82	& Lick	\\
Pr1\_ID02		&		01:04:15.8		&		+40:22:44.0		&		0.193		&		16.2		&	121.2	& 45.1	&	149.3	&	8.51	& Lick	\\
Pr1\_ID03		&		01:40:35.0		&		+23:44:51.0		&		0.320		&		16.7		&	186.7	& 45.0	&	155.3	&	8.49	& Lick	\\
Pr1\_ID04		&		02:27:39.7		&		+44:10:00.0		&		0.175		&		17.2		&	65.4	& 44.0	&	40.5	&	7.65	& Lick	\\
Pr1\_ID05		&		02:28:59.2		&		+39:08:44.0		&		0.336		&		17.1		&	165.1	& 44.3	&	67.7 	&	$^\dag$ & MDM	\\
Pr1\_ID06		&		03:37:02.9		&		+47:38:50.0		&		0.184		&		17.2		&	70.1	& 44.9	&	113.1	&	8.43	& Lick	\\
Pr1\_ID07		&		04:13:37.6		&		+72:06:52.7		&		0.105		&		15.3		&	88.9	& 44.1	&	44.2 	&	7.86	& MDM	\\
Pr1\_ID08		&		07:46:44.8		&		+29:40:59.0		&		0.292		&		17.2		&	128.0	& 44.5	&	81.3	&	7.98	& SDSS	\\
Pr1\_ID09		&		07:49:10.6		&		+28:42:14.6		&		0.344		&		17.3		&	156.2	& 44.8	&	119.3	&	8.81	& SDSS	\\
Pr1\_ID10		&		07:56:20.1		&		+30:45:35.4		&		0.236		&		15.2		&	258.9	& 44.8	&	105.7	&	8.44	& SDSS	\\
Pr1\_ID11		&		08:01:12.0		&		+51:28:12.3		&		0.321		&		17.3		&	140.9	& 44.2	&	59.7	&	7.99	& SDSS	\\
Pr1\_ID12		&		08:16:52.9		&		+24:16:12.6		&		0.276		&		17.3		&	113.2	& 44.6	&	90.8	&	8.30	& SDSS	\\
Pr1\_ID13		&		09:28:01.3		&		+49:18:17.3		&		0.115		&		16.7		&	49.8	& 43.7	&	25.3	&	7.29	& SDSS	\\
Pr1\_ID14		&		09:39:39.7		&		+37:57:05.8		&		0.231		&		17.2		&	94.3	& 44.5	&	72.0	&	8.24	& SDSS	\\
Pr1\_ID15		&		09:50:48.4		&		+39:26:50.5		&		0.206		&		16.4	    &	119.5	& 44.6	&	81.7	&	8.76	& SDSS	\\
Pr1\_ID16		&		10:05:28.3		&		+42:30:37.6		&		0.257		&		16.7		&	137.4	& 44.8	&	108.8	&	8.53	& SDSS	\\
Pr1\_ID17		&		10:26:13.9		&		+52:37:51.3		&		0.259		&		17.2		&	109.0	& 44.3	&	64.3	&	8.30	& SDSS	\\
Pr1\_ID18		&		10:59:35.5		&		+66:57:58.0		&		0.340		&		16.9		&	187.2	& 44.8	&	117.4	&	8.67	& SDSS	\\
Pr1\_ID19		&		11:04:13.9		&		+76:58:58.2		&		0.312		&		15.9		&	271.5	& 45.1	&	172.4	&	9.34	& SDSS	\\
Pr1\_ID20		&		11:05:27.3		&		+67:16:36.4		&		0.320		&		17.2		&	146.1	& 44.6	&	96.2	&	8.76	& SDSS	\\
Pr1\_ID21		&		11:15:06.0		&		+42:49:48.9		&		0.301		&		16.2		&	218.5	& 44.9	&	130.5	&	8.63	& Lick	\\
Pr1\_ID22		&		11:19:08.7		&		+21:19:18.0		&		0.176		&		15.2		&	179.1	& 45.1	&	147.1	&	8.71	& SDSS	\\
Pr1\_ID23		&		11:20:07.4		&		+42:35:51.4		&		0.226		&		16.8		&	111.5	& 44.4	&	70.8	&	8.77	& SDSS	\\
Pr1\_ID24		&		11:24:39.2		&		+42:01:45.1		&		0.225		&		16.0		&	161.0	& 44.9	&	119.9	&	8.36	& SDSS	\\
Pr1\_ID25		&		11:34:32.2		&		+60:46:34.8		&		0.201		&		17.0		&	86.5	& 44.2	&	48.6	&	8.51	& SDSS	\\
Pr1\_ID26		&		12:03:48.1		&		+45:59:51.1		&		0.343		&		16.8		&	197.1	& 44.9	&	143.3	&	8.88	& SDSS	\\
Pr1\_ID27		&		12:04:42.1		&		+27:54:11.8		&		0.165		&		15.0		&	177.4	& 44.5	&	72.4	&	9.02	& SDSS	\\
Pr1\_ID28		&		12:07:21.0		&		+26:24:29.2		&		0.324		&		16.9		&	174.0	& 45.0	&	155.8	&	8.49	& SDSS	\\
Pr1\_ID29		&		12:17:52.2		&		+33:34:47.3		&		0.178		&		17.3		&	64.0	& 44.4	&	64.0	&	8.64	& SDSS	\\
Pr1\_ID30		&		12:53:37.7		&		+21:26:18.2		&		0.127		&		15.3		&	111.7	& 44.4	&	63.5	&	8.77	& SDSS	\\
Pr1\_ID31		&		13:12:17.8		&		+35:15:21.1		&		0.183		&		15.5		&	163.3	& 44.7	&	97.5	&	8.57	& SDSS	\\
Pr1\_ID32		&		13:56:32.8		&		+21:03:52.4		&		0.300		&		15.9		&	259.3	& 45.1	&	166.7	&	8.78	& SDSS	\\
Pr1\_ID33		&		14:03:08.8		&		+37:58:27.5		&		0.184		&		16.0		&	126.3	& 44.5	&	69.2	&	8.23	& SDSS	\\
Pr1\_ID34		&		14:08:39.0		&		+63:06:00.5		&		0.261		&		16.7		&	141.3	& 44.7	&	103.9	&	8.67	& SDSS	\\
Pr1\_ID35		&		14:08:54.2		&		+56:57:43.4		&		0.336		&		17.3		&	149.7	& 44.9	&	137.4	&	8.79	& SDSS	\\
Pr1\_ID36		&		14:15:35.9		&		+48:35:43.6		&		0.275		&		17.0		&	131.1	& 44.5	&	83.2	&	8.46	& SDSS	\\
Pr1\_ID37		&		14:46:45.9		&		+40:35:05.8		&		0.267		&		16.0		&	210.2	& 45.1	&	164.5	&	8.44	& SDSS	\\
Pr1\_ID38		&		14:56:08.6		&		+38:00:38.6		&		0.283		&		16.8		&	149.1	& 44.6	&	84.8	&	9.09	& SDSS	\\
Pr1\_ID39		&		15:15:35.3		&		+48:05:30.6		&		0.312		&		15.9		&	268.2	& 45.1	&	178.9	&	8.75	& SDSS	\\
Pr1\_ID40		&		15:26:24.0		&		+27:54:52.1		&		0.231		&		17.1		&	99.0	& 44.7	&	94.3	&	8.58	& SDSS	\\
Pr1\_ID41		&		15:40:04.3		&		+35:50:50.1		&		0.164		&		17.2		&	60.2	& 44.2	&	51.3	&	8.08	& SDSS	\\
Pr1\_ID42		&		15:47:43.5		&		+20:52:16.8		&		0.264		&		16.1		&	197.2	& 45.0	&	139.9	&	9.11	& SDSS	\\
Pr1\_ID43		&		16:19:11.2		&		+50:11:09.2		&		0.234		&		16.8		&	116.4	& 44.4	&	63.8	&	8.69	& SDSS	\\
Pr1\_ID44		&		16:46:49.1		&		+38:25:04.6		&		0.311		&		17.2		&	141.6	& 44.7	&	102.9	&	7.93	& SDSS	\\
Pr1\_ID45		&		17:12:06.3		&		+58:42:17.0		&		0.310		&		17.1		&	146.8	& 44.8	&	121.0 	&	$^\dag$ & Lick	\\
Pr1\_ID46		&		17:14:48.5		&		+33:27:38.3		&		0.181		&		17.2		&	68.2	& 44.4	&	64.9	&	8.84    & SDSS	\\
Pr1\_ID47		&		19:35:21.2		&		+53:14:12.1		&		0.248		&		17.0		&	113.5	& 45.0	&	146.9	&	8.64    & Lick	\\
Pr1\_ID48		&		23:03:29.9		&		+45:37:40.6		&		0.301		&		17.0		&	147.6	& 44.7	&	99.6 	&	8.91    & MDM	\\
\tableline
\end{tabular}
\tablecomments{
Columns: (1) Target ID. 
(2) RA in J2000.0. 
(3) Dec in J2000.0
(4) redshift. 
(5) B-band magnitude from literature.
(6) (1+z)lag values in days estimated from B-band magnitude in col.(5).
(7) Measured L$_{5100}$ from a reference spectrum.
(8) Measured (1+z)lag value using col.(7) and Eq.(3).
(9) Black hole mass estimated using Eq.(4) with the velocity dispersion($\sigma$) of a broad H$\beta$ emission line and L$_{5100}$ in col. (7). 
$^\dag$ Pr1\_ID05 and Pr1\_ID45 - No H$\beta$ line was detected. 
(10) Reference of the individual spectrum: SDSS - SDSS DR12 archive, BG92 - Boroson \& Green (1992), Lick \& MDM  - obtained with our program.
}
\end{center}
\end{table*}

\begin{table*}
\begin{center}
\caption{Properties of our samples\label{tbl_target}}
\begin{tabular}{lllccrcrrr}
\tableline\tableline
  Name & RA & Dec & z & B & (1+z)lag & 	log L$_{5100,spec}$	& (1+z)lag$_{spec}$  & logM$_{\rm BH}$ & Ref.Spec. \\
      &        &          &     &       & (day) & (erg s$^{-1}$)	& (day) & (M$_{\odot}$) &  \\
 (1)  & (2)    & (3)      & (4) & (5)   & (6)   & (7)   			& (8)  	& (9) & (10)	\\
\tableline
Pr2\_ID01	&	08:03:08.6	&	+53:00:04.8	&	0.287	&	17.3	&	119.9	&	44.3	&	58.9	&	8.24 & SDSS	\\
Pr2\_ID02	&	08:13:17.9	&	+43:56:20.6	&	0.254	&	17.0	&	116.9	&	44.4	&	65.6	&	8.80 & SDSS	\\
Pr2\_ID03	&	08:44:45.3	&	+76:53:09.6	&	0.131	&	16.3	&	71.2	&	44.1	&	44.1	&	7.99 & SDSS	\\
Pr2\_ID04	&	08:48:53.1	&	+28:24:11.8	&	0.198	&	16.5	&	107.6	&	44.0	&	41.8	&	8.52 & SDSS	\\
Pr2\_ID05	&	09:04:27.3	&	+37:43:57.5	&	0.198	&	17.3	&	72.7	&	44.2	&	50.3	&	7.79 & SDSS	\\
Pr2\_ID06	&	09:36:08.7	&	+65:10:54.9	&	0.192	&	17.0	&	81.7	&	44.3	&	60.9	&	7.78 & SDSS	\\
Pr2\_ID07	&	09:37:02.9	&	+68:24:08.4	&	0.295	&	15.8	&	258.3	&	44.2	&	55.0	&	8.35 & SDSS	\\
Pr2\_ID08	&	10:56:09.8	&	+55:16:04.0	&	0.256	&	16.3	&	167.1	&	44.3	&	58.2	&	8.21 & SDSS	\\
Pr2\_ID09	&	11:17:06.4	&	+44:13:33.3	&	0.144	&	16.1	&	89.8	&	44.3	&	56.3	&	8.58 & SDSS	\\
Pr2\_ID10	&	11:18:30.3	&	+40:25:54.0	&	0.154	&	16.0	&	99.9	&	44.2	&	47.8	&	7.93 & SDSS	\\
Pr2\_ID11	&	11:36:17.1	&	+44:10:22.6	&	0.198	&	17.2	&	76.3	&	44.2	&	50.3	&	8.60 & SDSS	\\
Pr2\_ID12	&	11:52:53.5	&	+45:34:02.9	&	0.211	&	17.2	&	83.0	&	44.2	&	53.4	&	8.76 & SDSS	\\
Pr2\_ID13	&	12:20:28.1	&	+40:50:35.0	&	0.222	&	16.3	&	138.6	&	44.3	&	62.4	&	8.81 & SDSS	\\
Pr2\_ID14	&	12:28:30.9	&	+28:14:11.8	&	0.100	&	13.2	&	233.3	&	44.0	&	38.9	&	8.59 & SDSS	\\
Pr2\_ID15	&	12:59:08.4	&	+56:15:30.7	&	0.161	&	17.2	&	58.6	&	44.2	&	51.2	&	8.80 & SDSS	\\
Pr2\_ID16	&	13:05:16.8	&	+26:13:04.0	&	0.183	&	17.2	&	69.2	&	44.2	&	47.9	&	8.11 & SDSS	\\
Pr2\_ID17	&	13:20:59.4	&	+29:57:28.1	&	0.206	&	17.2	&	80.7	&	44.4	&	67.1	&	8.80 & SDSS	\\
Pr2\_ID18	&	13:23:49.5	&	+65:41:48.2	&	0.168	&	15.9	&	120.1	&	44.4	&	67.4	&	8.30 & SDSS	\\
Pr2\_ID19	&	13:30:16.1	&	+52:51:01.9	&	0.162	&	16.4	&	87.9	&	44.3	&	56.5	&	8.91 & SDSS	\\
Pr2\_ID20	&	14:05:02.6	&	+47:07:47.5	&	0.152	&	14.9	&	169.1	&	44.1	&	42.8	&	8.10 & SDSS	\\
Pr2\_ID21	&	14:05:16.2	&	+25:55:34.2	&	0.164	&	15.6	&	133.6	&	44.3	&	54.6	&	8.04 & SDSS	\\
Pr2\_ID22	&	14:06:21.9	&	+22:23:46.6	&	0.098	&	15.8	&	63.4	&	44.2	&	45.0	&	7.58 & SDSS	\\
Pr2\_ID23	&	14:17:00.8	&	+44:56:06.6	&	0.114	&	15.7	&	78.3	&	44.2	&	46.8	&	7.97 & SDSS	\\
Pr2\_ID24	&	14:29:43.1	&	+47:47:26.2	&	0.220	&	16.3	&	135.0	&	44.4	&	65.4	&	8.21 & SDSS	\\
Pr2\_ID25	&	14:37:47.9	&	+28:30:19.5	&	0.249	&	17.2	&	104.2	&	44.1	&	48.7	&	8.58 & SDSS	\\
Pr2\_ID26	&	14:42:07.5	&	+35:26:23.0	&	0.079	&	15.0	&	73.1	&	44.2	&	48.7	&	7.69 & SDSS	\\
Pr2\_ID27	&	14:53:45.7	&	+34:33:20.3	&	0.209	&	17.2	&	81.9	&	44.2	&	51.4	&	8.75 & SDSS	\\
Pr2\_ID28	&	15:21:14.3	&	+22:27:43.9	&	0.137	&	16.1	&	82.9	&	44.1	&	42.2	&	7.82 & SDSS	\\
Pr2\_ID29	&	15:26:21.7	&	+43:23:49.6	&	0.156	&	16.3	&	88.5	&	44.0	&	39.4	&	7.82 & SDSS	\\
Pr2\_ID30	&	15:33:54.7	&	+23:56:14.8	&	0.232	&	17.2	&	94.3	&	44.2	&	52.3	&	8.72 & SDSS	\\
Pr2\_ID31	&	15:35:39.2	&	+56:44:06.5	&	0.207	&	16.8	&	98.3	&	44.3	&	60.2	&	9.08 & SDSS	\\
Pr2\_ID32	&	15:42:34.8	&	+57:41:41.9	&	0.245	&	17.3	&	96.5	&	44.4	&	66.8	&	7.98 & SDSS	\\
Pr2\_ID33	&	15:44:50.0	&	+43:51:51.3	&	0.149	&	16.9	&	61.6	&	44.1	&	41.1	&	8.76 & SDSS	\\
Pr2\_ID34	&	16:05:07.9	&	+48:34:22.1	&	0.295	&	16.7	&	166.1	&	44.9	&	131.5	&	8.83 & SDSS	\\
Pr2\_ID35	&	16:14:13.2	&	+26:04:16.3	&	0.131	&	16.0	&	82.5	&	44.4	&	62.2	&	8.26 & SDSS	\\
Pr2\_ID36	&	16:25:33.8	&	+37:10:42.8	&	0.202	&	17.0	&	86.6	&	44.2	&	50.4	&	7.60 & SDSS \\
Pr2\_ID37	&	16:27:56.1	&	+55:22:31.5	&	0.133	&	16.2	&	76.7	&	44.4	&	63.8	&	8.45 & BG92	\\
\tableline
\end{tabular}
\tablecomments{
Columns: (1) Target ID. 
(2) RA in J2000.0. 
(3) Dec in J2000.0
(4) redshift. 
(5) B-band magnitude from literature.
(6) (1+z)lag values in days estimated from B-band magnitude in col.(5).
(7) Measured L$_{5100}$ from a reference spectrum.
(8) Measured (1+z)lag value using col.(7) and Eq.(3).
(9) Black hole mass estimated using Eq.(4) with the velocity dispersion($\sigma$) of a broad H$\beta$ emission line and L$_{5100}$ in col. (7). 
(10) Reference of the individual spectrum: SDSS - SDSS DR12 archive, BG92 - Boroson \& Green (1992), Lick \& MDM  - obtained with our program.
}
\end{center}
\end{table*}

\begin{table*}
\begin{center}
\caption{Properties of our samples\label{tbl_target}}
\begin{tabular}{lllccrcrrr}
 \tableline\tableline
  Name & RA & Dec & z & B & (1+z)lag & 	log L$_{5100,spec}$	& (1+z)lag$_{spec}$  & logM$_{\rm BH}$ & Ref.Spec. \\
      &        &          &     &       & (day) & (erg s$^{-1}$)	& (day) & (M$_{\odot}$) &  \\
 (1)  & (2)    & (3)      & (4) & (5)   & (6)   & (7)   			& (8)  	& (9) & (10)	\\
\tableline
P01	&	00:05:59.2	&	+16:09:49.0	&	0.451	&	16.0	&	455.8	&	45.6	&	331.2	&	9.48 & BG92 \\
P02	&	00:10:31.0	&	+10:58:29.5	&	0.089	&	16.1	&	49.2	&	44.4	&	57.0	&	8.74 & BG92 \\
P03	&	00:29:13.7	&	+13:16:04.0	&	0.142	&	15.0	&	159.6	&	44.6	&	79.3	&	8.61 & BG92 \\
P04	&	00:45:47.2	&	+04:10:23.3	&	0.385	&	15.9	&	370.9	&	45.2	&	213.5	&	8.81 & BG92 \\
P05	&	00:53:34.9	&	+12:41:35.9	&	0.059	&	14.4	&	70.0	&	44.3	&	49.6	&	7.92 & BG92 \\
P06	&	00:54:52.1	&	+25:25:39.0	&	0.154	&	15.4	&	142.6	&	44.7	&	90.6	&	9.06 & BG92 \\
P07	&	01:59:50.3	&	+00:23:40.8	&	0.163	&	15.2	&	160.5	&	44.6	&	76.9	&	9.23 & BG92 \\
P08	&	21:14:52.6	&	+06:07:42.5	&	0.466	&	15.5	&	595.8	&	45.6	&	338.8	&	9.29 & BG92 \\
P09	&	21:32:27.8	&	+10:08:19.2	&	0.063	&	14.9	&	66.9	&	44.3	&	50.4	&	8.14 & BG92 \\
P10	&	22:11:53.9	&	+18:41:50.0	&	0.070	&	15.9	&	41.7	&	43.6	&	23.1	&	8.44 & BG92 \\
P11	&	22:17:12.3	&	+14:14:20.9	&	0.066	&	15.0	&	60.1	&	44.4	&	55.1	&	8.35 & BG92 \\
P12	&	22:36:07.7	&	+13:43:55.4	&	0.326	&	16.0	&	267.7	&	45.0	&	148.6	&	8.32 & BG92 \\
P13	&	22:54:10.4	&	+11:36:38.7	&	0.326	&	16.3	&	239.7	&	45.5	&	281.2	&	9.23 & BG92 \\
P14	&	23:11:17.8	&	+10:08:15.8	&	0.433	&	16.1	&	398.1	&	45.2	&	218.2	&	9.50 & BG92 \\
P15	&	23:51:56.1	&	-01:09:13.3	&	0.174	&	15.0	&	190.1	&	44.6	&	84.6	&	8.98 & Lick \\
\tableline
\end{tabular}
\tablecomments{
Columns: (1) Target ID. 
(2) RA in J2000.0. 
(3) Dec in J2000.0
(4) redshift. 
(5) B-band magnitude from literature.
(6) (1+z)lag values in days estimated from B-band magnitude in col.(5).
(7) Measured L$_{5100}$ from a reference spectrum.
(8) Measured (1+z)lag value using col.(7) and Eq.(3).
(9) Black hole mass estimated using Eq.(4) with the velocity dispersion($\sigma$) of a broad H$\beta$ emission line and L$_{5100}$ in col. (7). 
(10) Reference of the individual spectrum: SDSS - SDSS DR12 archive, BG92 - Boroson \& Green (1992), Lick \& MDM  - obtained with our program.
}
\end{center}
\end{table*}

\section{sample}

\begin{figure}
\includegraphics[height=5cm, width = 0.23\textwidth]{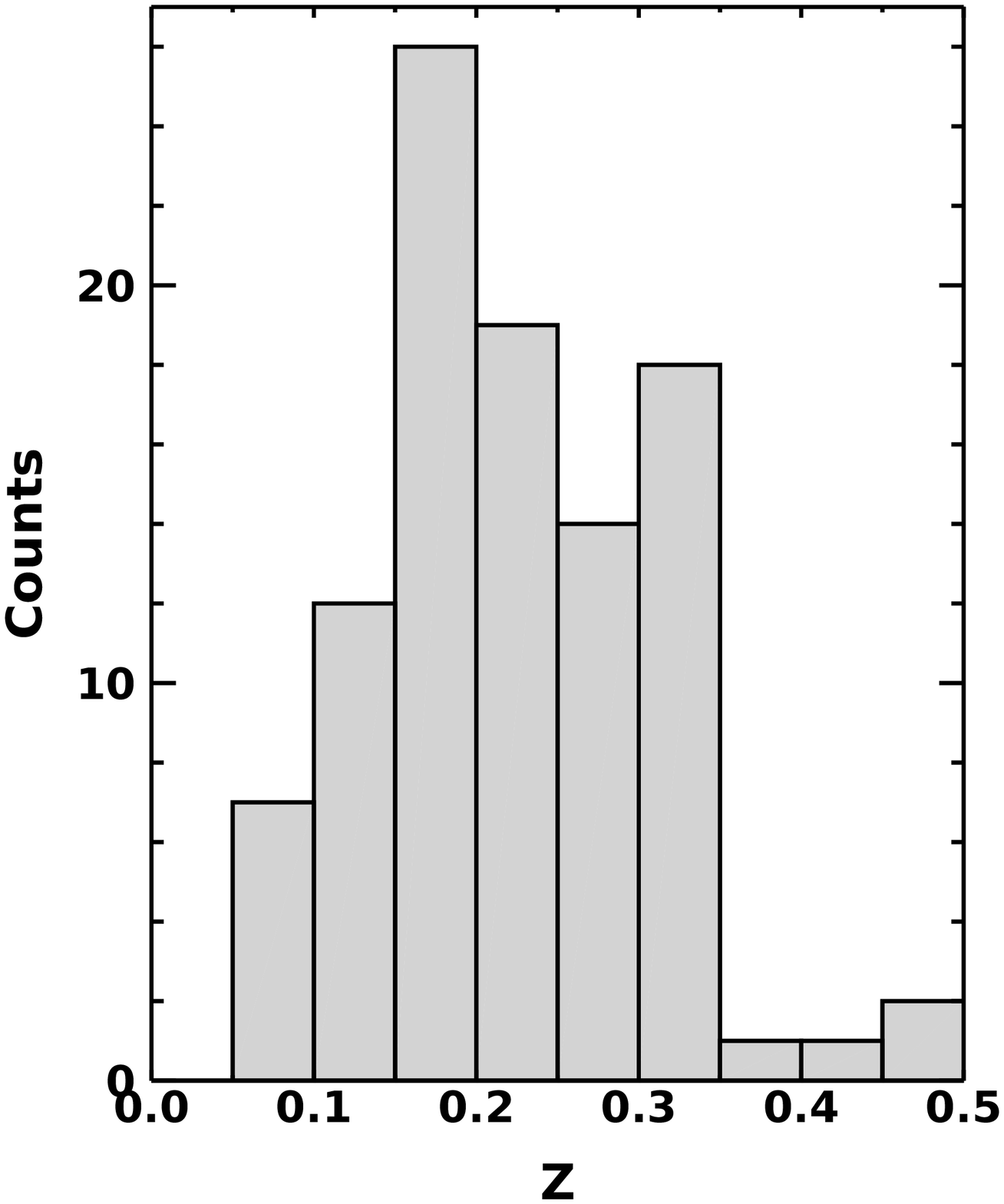}
\includegraphics[height=5cm, width = 0.23\textwidth]{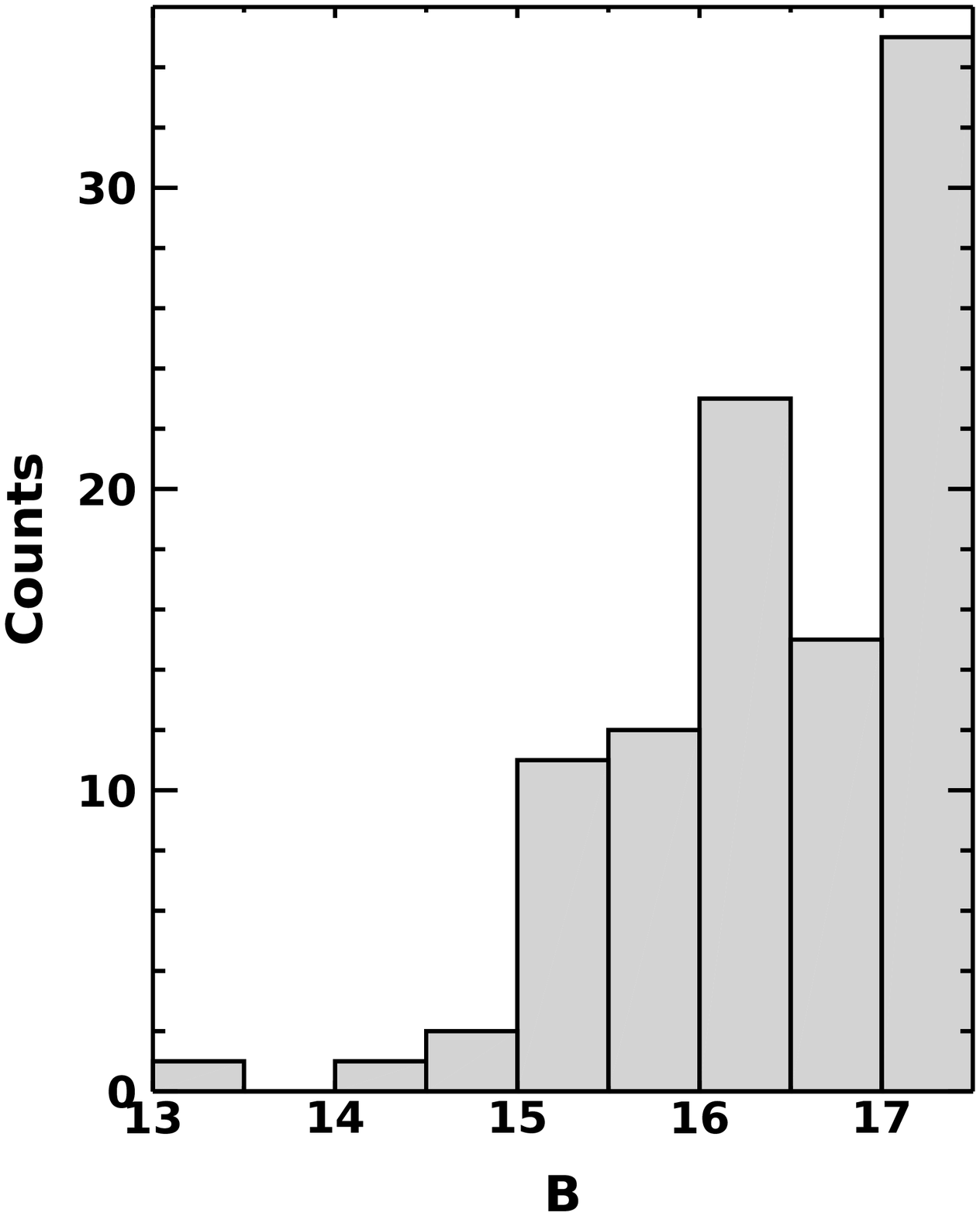}\\
\caption{Distributions of the redshift (left) and the B-band magnitude (right) of the sample. 
}
\end{figure}

\begin{figure}
\includegraphics[width = 0.35\textwidth]{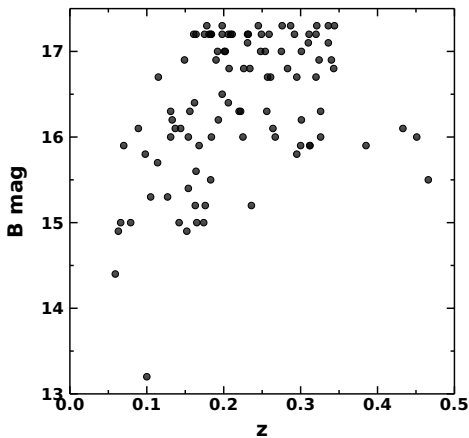}
\caption{Comparison of the B-band magnitude and redshift of the sample}
\end{figure}

\begin{figure}
\includegraphics[width = 0.35\textwidth]{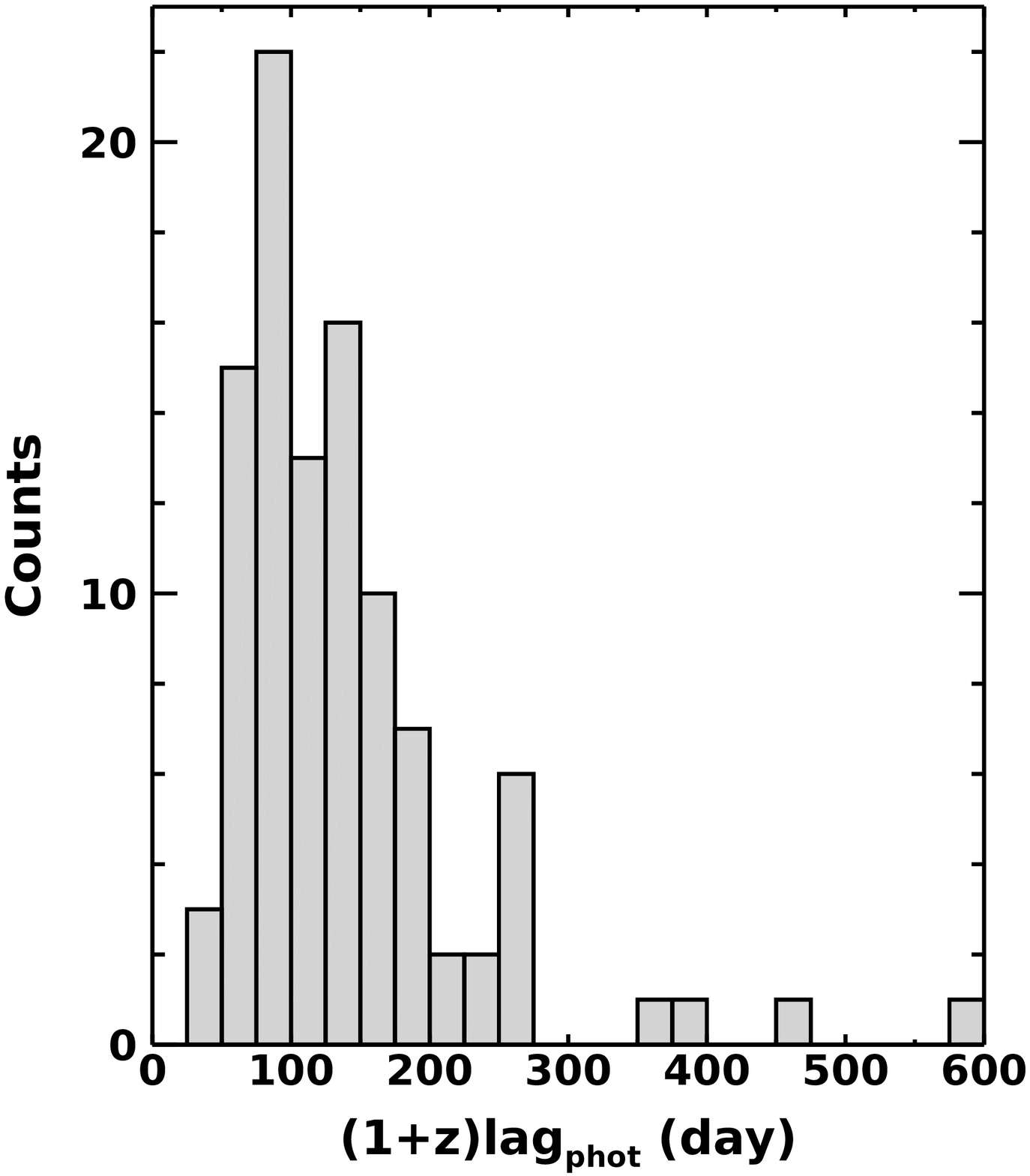}
\caption{Distribution of the expected lag calculated from the monochromatic luminosity at 5100\AA, which is estimated based on broad-band photometry
}
\end{figure}
\subsection{Sample selection}\label{sel}

We use the  MILLIQUAS Catalog to select AGNs at z $<$ 0.5 and with the V-band magnitude $<$ 17. 
The expected lags are estimated from the optical luminosity at 5100\AA, based on the size-luminosity relation \citep{Bentz+13}. 
Using available spectra from various archives, we perform spectral decomposition to separate the power-law continuum, Fe II emission, and stellar component. Then, we measure the rest-frame luminosity at 5100\AA\ (L$_{5100}$) from the power-law component.
 For 11 AGNs, for which any archival spectrum is not available, we use the spectrum obtained at the Lick 3-m or MDM 2.4-m telescopes 
during our initial spectroscopy test. Details of our spectroscopic observations are summarized by \citet{Rakshit+19}.

As a primary sample, we select 48 AGNs at Dec $>$20 degree and the expected lag 70 $<$ (1+z)lag $<$ 250 day.
Considering the large uncertainties in the estimated lag, we also select 37 AGNs with a shorter lag 40 $<$ (1+z)lag $<$ 70 day. 
We call these two groups as primary 1 (Pr1) and primary 2 (Pr2) groups as presented in Table 3 and 4, respectively. 
In addition, we choose a sample of 15 AGNs from the Palomar Green quasar catalogue by lowering the declination limit (i.e., Dec $>$ 0 degree), in order to increase the sample size. 
Thus, a total of 100 AGNs at z $<$ 0.5 is selected for initial imaging and spectroscopy in order to examine the number of comparison stars in the field of view and the shape of the \Hb\ and [OIII] emission lines, for a feasibility test for differential photometry and spectral decomposition. 

\begin{figure}
\includegraphics[width = 0.4\textwidth]{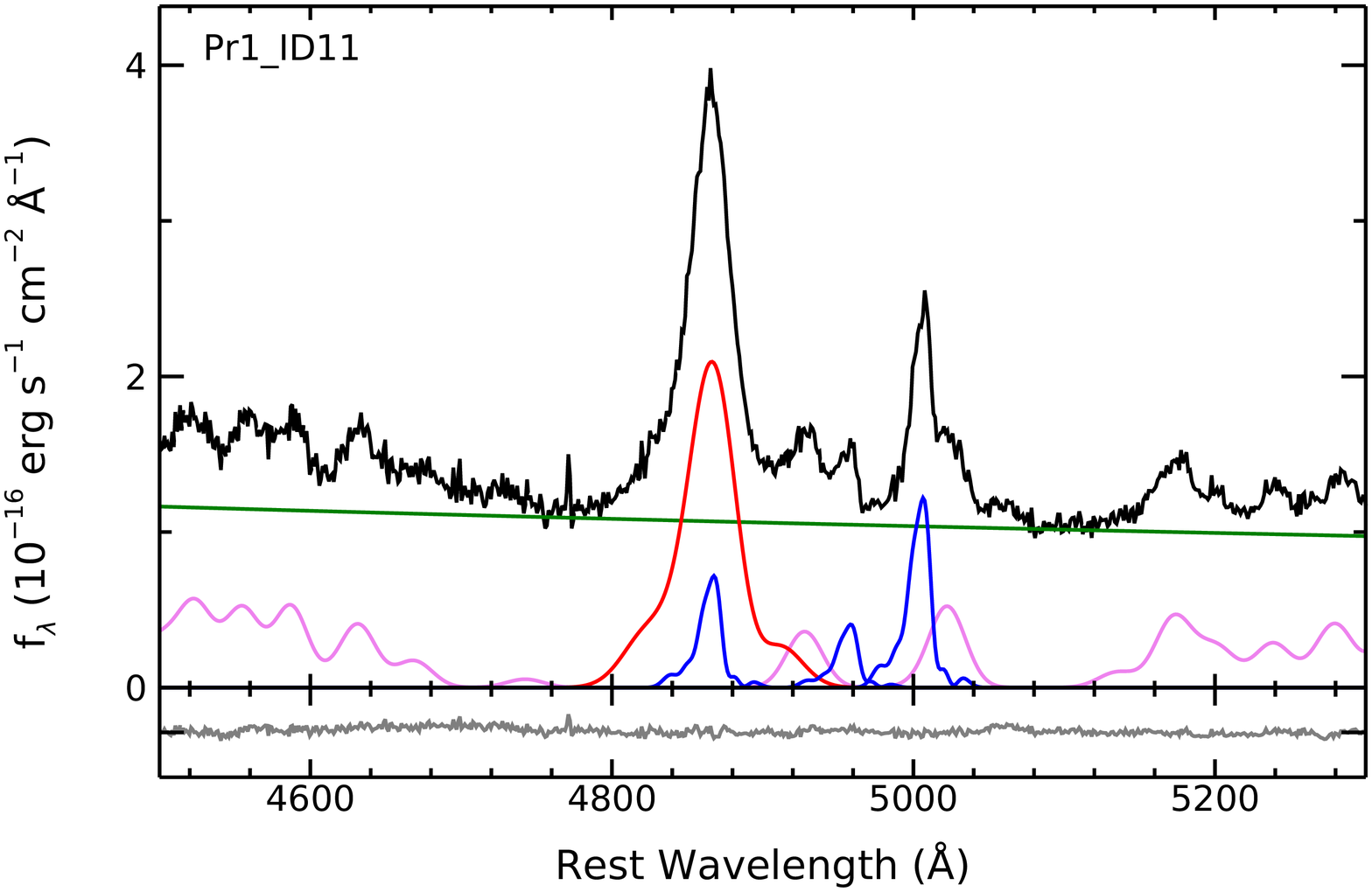}
\includegraphics[width = 0.4\textwidth]{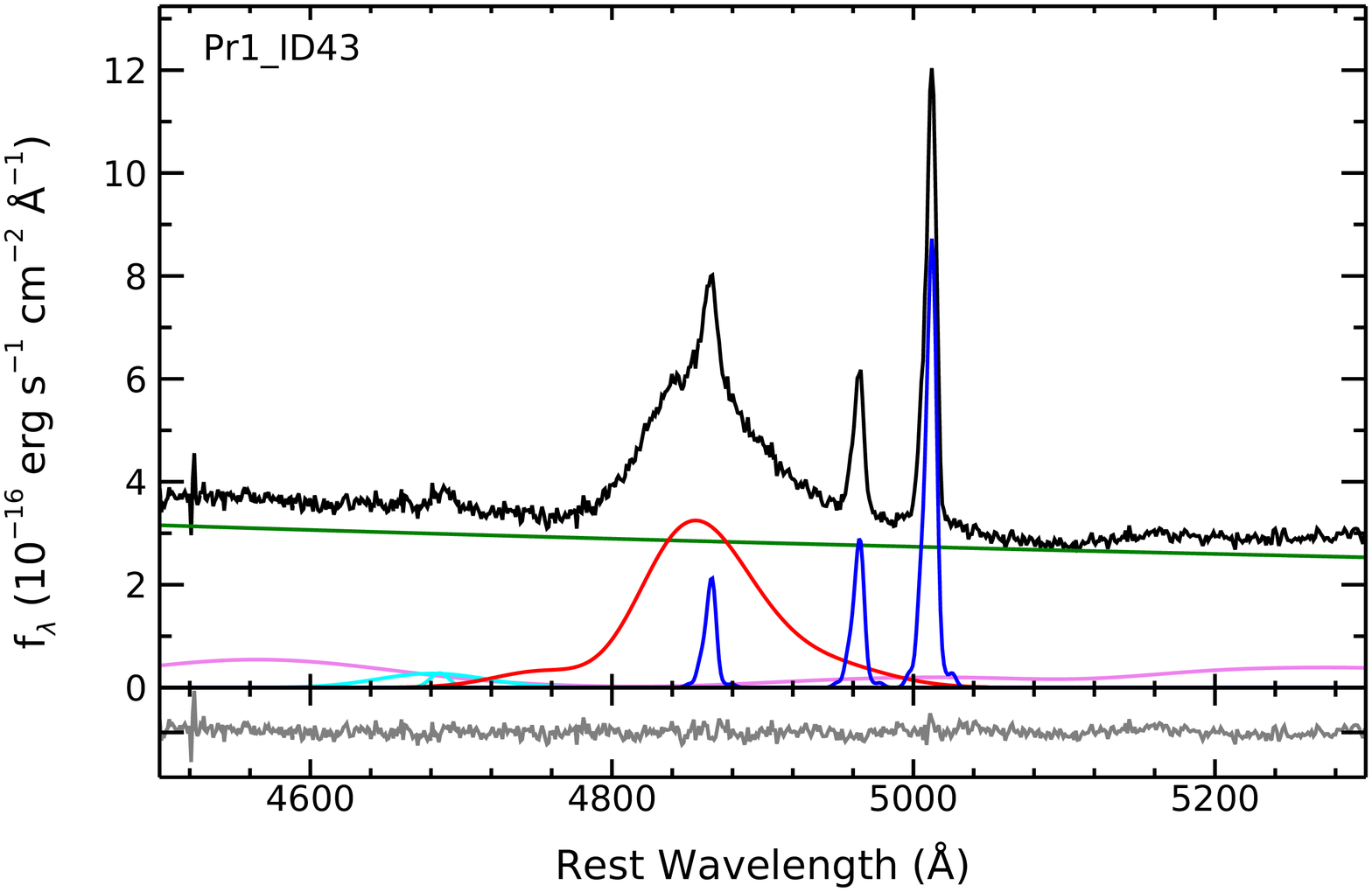}
\includegraphics[width = 0.4\textwidth]{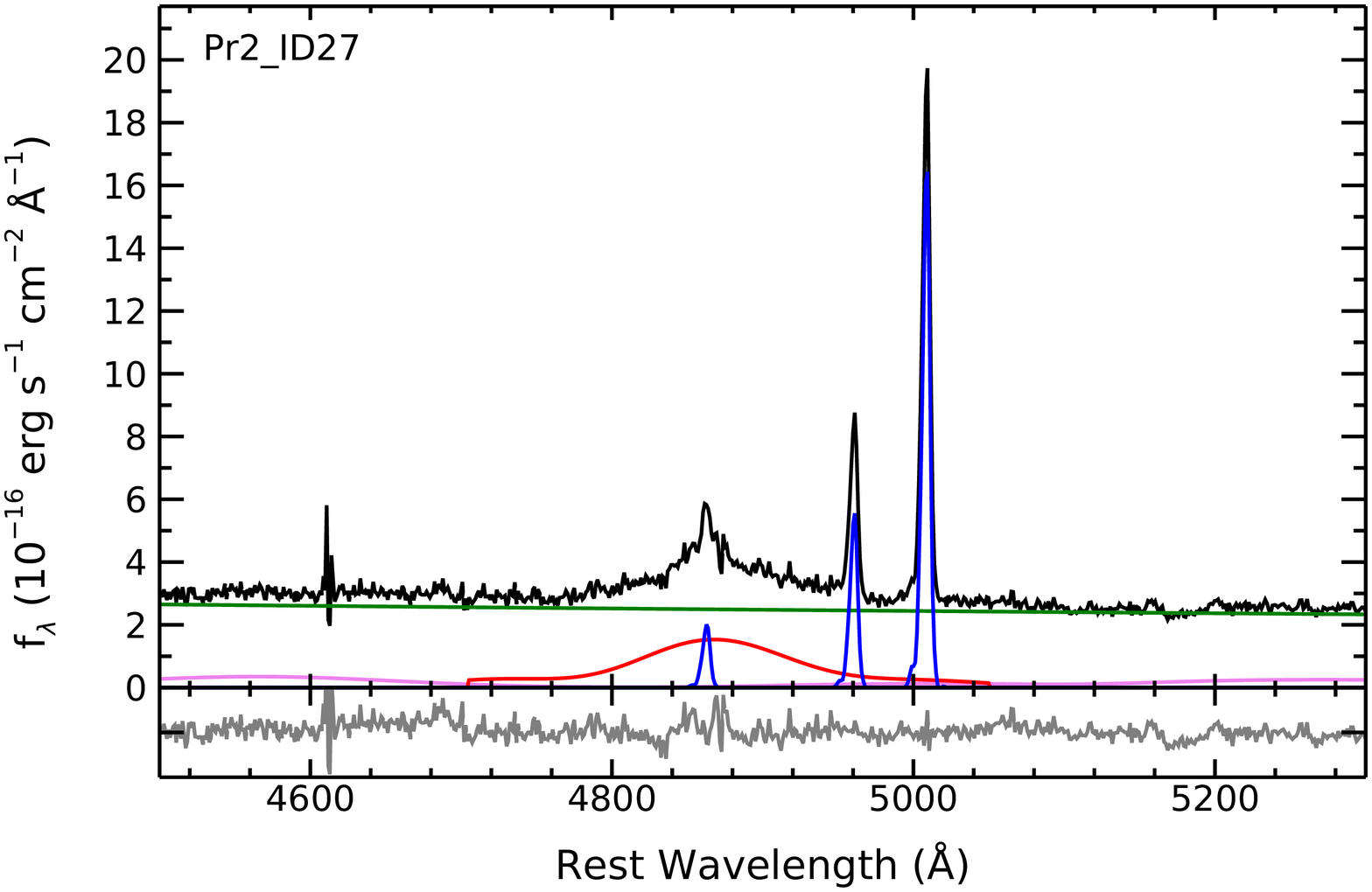}
\caption{Examples of spectral decomposition for three targets. The observed spectrum (black) is fitted with three components: a power-law (green), Fe II emission blends (magenta), and a stellar population model. Note that since stellar flux is weak, the stellar component is negligible in the fit (no stellar component). The broad \Hb\ (red) as well as narrow \Hb\ and OIII lines (blue) are separately fitted. 
}
\end{figure}

\subsection{Sample properties}

In Figure 5, we present the redshift distribution of the sample with a mean redshift is $0.22\pm0.09$. 
We limit the redshift as z$\sim$0.5 in order to properly locate the \Hb\ emission line within the spectral range of the optical spectral setup. Thus, we can avoid the change of the grating or tilt angle of the spectrograph during the observing runs, to minimize the overhead and systematic uncertainties. For high-z AGNs, the (1+z) factor becomes significantly large, increasing the time lag in the observed-frame by over 50\% at z $>$ 0.5. 

The distribution of the B-band magnitude is presented in Figure 5 (right panel), ranging from B=13.2 to B=17.3, with a mean $<B> = 16.3\pm0.8$. Note that the B-band magnitude is obtained from various available catalogues in the literature. Thus, we expect that the current B-band magnitude will be somewhat different due to the variability over time. We compare the distributions of the redshift and B-band magnitude of each target in Figure 6, demonstrating the difficulty of the time lag measurements for higher-z targets, due to the average fainter magnitude as well as the (1+z) factor. For the continuum photometry monitoring, we will use the B-band filter for AGNs at z$<$0.3, while we will use the V-band filter for higher-z AGNs, by considering the location of the \Hb\ emission line in the observed spectra. Thus, with the choice of the photometry filter, we can avoid the variable \Hb\ emission line in the broad-band photometry, although the choice of the filter does not perfectly avoid the emission line, depending on the redshift of each target. 

To estimate the \Hb\ lag of each target, we calculate the optical luminosity using B-band magnitude.
First, we calculate the absolute B-band magnitude ($M_B$) using an equation in Veron-Cetty \& Veron (2010) assuming an optical spectral index $\alpha = 0.3$ as follows:
\begin{equation}
M_{B} = m_{B} + 5 - 5\times logD - k + \Delta m(z)
\end{equation}
where $D$ is the luminosity distance, $z$ is the redshift, $k$ is the k-correction factor ($= -2.5~ log(1+z)^{1-\alpha}$), and $\Delta$m(z) is
a correction to $k$ considering that the spectrum is not a power law (see Table 2 in their paper for the values of $\Delta$m(z) for $z < 5.0$). 
We then convert the absolute magnitude $M_B$ into the luminosity at 5100\AA~ using Eq. 2 in Marziani et al. (2003), assuming galactic extinction $A_{B} = 0$.
\begin{equation} 
L_{5100} = 3.137 \times 10^{35-0.4 M_{B}} ~(\rm {erg~ s^{-1}})
\end{equation}
Using the estimated $L_{5100}$ from the B-band magnitude, we estimate the \Hb\ lag, using the size-luminosity relation \ from \citet{Bentz+13}:
\begin{equation}
log \tau~ (\rm light-day) = 1.527 +0.533~ log~ (L_{5100} /(10^{44} erg s^{-1}))
\end{equation}
The expected lag ($\tau$) is presented in the Table 1--3. 
In Figure  7, we show the distribution of the expected \Hb\ lag, ranging from 42 to 596 light days with a mean $140\pm87$ day.  As expected, the majority of targets has \Hb\ lag around 100-300 days, while a small fraction of the sample has a larger lag over 300 days. Note that the estimated lags suffer large uncertainty due to the transformation from broad-band photometry.

In order to measure AGN properties and estimate the \Hb\ lag from the luminosity at 5100\AA, we analyze the spectrum of each target. The raw spectra have been compiled from the SDSS DR12 archive for targets listed in the MILLIQUAS Catalog while we also used NED for the PG quasars \citep{BG92}. 
We perform spectral decomposition to separate AGN continuum
by following the procedure given by previous studies \citep{Woo+06, Park+12, Woo+15}.
Here, we briefly describe the procedure for completeness.
We use three main components to fit the continuum: a power-law component representing featureless AGN continuum, a Fe II template from \citet{BG92} to fit the Fe II emission blends, and the stellar population model to represent host galaxy star light, using the spectral ranges, 4430$-$4600\AA\ and 5080$-$5550\AA, where Fe II emission blends dominate. For the multicomponent analysis, we use the nonlinear Levenberg-Marquardt least-squares fitting routine $mpfit$ \citep{Markwardt2009} in IDL. In Figure 8, we demonstrate the fitting procedure for three examples. Based on the best-fit results, we also determine the monochromatic luminosity at 5100\AA~ from the power-law component. 

In Figure 9, we present the estimated \Hb\ lag based on the measured L$_{5100}$. The mean lag of the sample is 90 light day with a standard deviation of 58 light day. The calculated \Hb\ lag is somewhat different compared to that estimated from photometry (see Figure 10), reflecting the uncertainty in the transformation of the broad band magnitude to the monochromatic luminosity at 5100\AA, and the variability as the photometry and spectra were not simultaneous. Since L$_{5100}$ is a directly measured monochromatic luminosity at 5100\AA, the \Hb\ lag based on spectroscopy is more appropriate to use for designing our monitoring campaign. Even for the spectroscopy-based \Hb\ lag, there are a number of uncertainties, including the flux uncertainty from spectrophotometric calibration, systematic uncertainty in separating AGN continuum from stellar component and Fe II emission, and the intrinsic scatter in the size-luminosity relation. Moreover, as the variability of AGN will change the optical luminosity and the \Hb\ lag as a function of time, the estimated lag should be used as a guidance of the sample, and the true \Hb\ lag should be measured based on the cross correlation analysis.

We determine the \mbh\ of each target using the measured optical luminosity at 5100\AA\ (L$_{5100}$) and the line dispersion of the \Hb\ line, based 
on the \Hb-based recipe,
\begin{equation}
M_{\rm BH} = f \times 10^{6.819} 
\left(\frac{\rm \sigma_{H\beta}}{10^3~\rm {km~s^{-1}}}\right)^{2}
\left(\frac{\lambda L_{5100}}{10^{44}~{\rm erg~s^{-1}}}\right)^{0.533}~{\rm M}_{\odot}~.
\end{equation}
where the {\it f} is the viral factor, which is empirically determined based on the \mbh-stellar velocity dispersion relation by \cite{Woo+15}. 
While we will measure the \Hb\ lag and determine the reverberation-based \mbh, the estimated mass enables us to investigate the 
properties of the sample although there is large uncertainties. In Figure 11, we compare the derived \mbh\ with bolometric AGN luminosity, which is calculated by multiplying a factor of 10 to L$_{5100}$ \citep{Woo+02}. The \mbh\ of the sample ranges from log \mbh =  7.3 to 9.5 and the Eddington
ratio ranges from 1\% to close to 100\%, indicating that the sample contains intermediate to high Eddington AGNs.

\begin{figure}
\includegraphics[width = 0.45\textwidth]{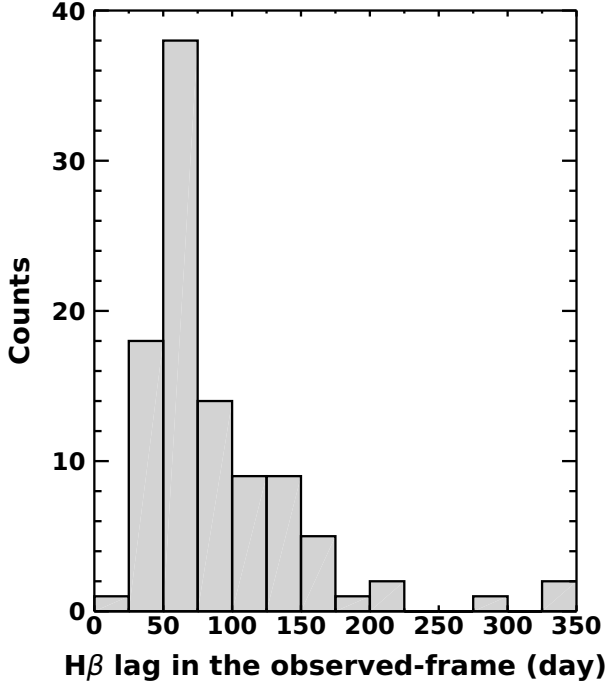}
\caption{Distribution of the expected lag calculated from the monochromatic luminosity at 5100\AA\ based on the spectral decomposition analysis}
\end{figure}

\begin{figure}
\includegraphics[width = 0.45\textwidth]{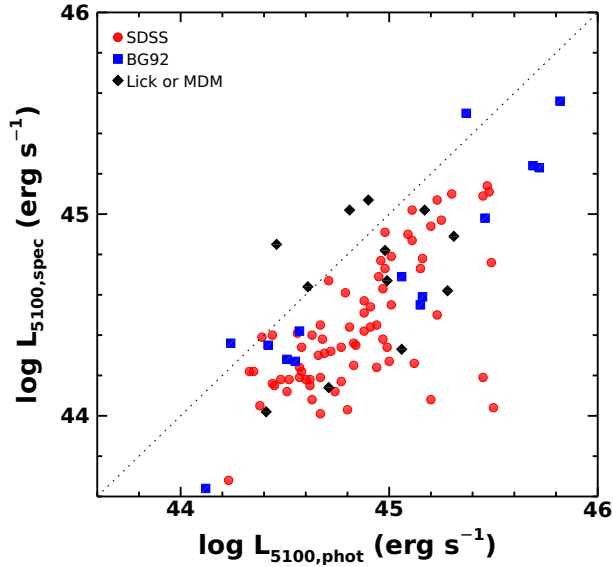}
\caption{ Comparison of the luminosity at 5100\AA\ between the estimates based on the broad-band photometry and the measurements from the spectral decomposition analysis. Each symbol represents the reference spectrum: 
SDSS spectrum (red dot), Boroson \& Green (1992) (blue square), Lick 3-m or MDM 2.4-m (black diamond). 
}
\end{figure}

\begin{figure}
\includegraphics[width = 0.45\textwidth]{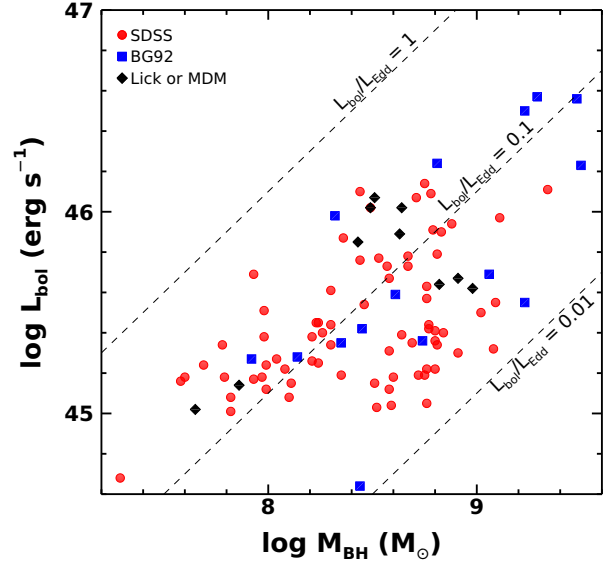}
\caption{Distributions of \mbh\ and bolometric luminosity. The calculated Eddington ratios are denoted with dashed lines. 
Symbols are same as in Fig. 10.
}
\end{figure}

\section{Summary and Conclusions}\label{section:sum}

To overcome the current limitations of the reverberation studies, which have mainly focused on relatively low-to-moderate luminosity AGNs
at low redshift, we developed a new reverberation mapping project, aiming at measuring \Hb\ lag for relatively luminous AGNs
out to z$\sim$0.5. We described the strategy of the Seoul National University AGN Monitoring Project, which was designed to be carried
out over 3 years. We performed simulations to generate light curves, measured the success rate for the lag measurements using various 
sets of parameters, including campaign duration, time cadence, seasonal gap, and variability characteristics. 
Based on the simulation results and the availability of telescope time, we determined the optimal cadence and sample parameters for 
a three year monitoring project. 

While reverberation mapping for high-z AGNs is much more challenging due to the large (1+z) factor, requiring a factor of (1+z) longer
monitoring campaign, the recent CIV lag measurements from a long-term campaign clearly showed the feasibility \citep{Lira+18}. However, in order to study the \Hb\ size-luminosity relation with very luminous AGNs at high z, a monitoring campaign with infra-red spectroscopy is required, due to the redshift of the \Hb\ emission line, which is currently much more challenging.

We selected a sample of 100 AGNs for the project, and described the photometric and spectroscopic properties using the archival data. 
The expected lag based on the monochromatic luminosity at 5100\AA\ ranges between 40 and $\sim$350 light days, which 
can be measured with a three year monitoring project. As the sample covers high luminosity and high Eddington ratio AGNs at higher redshift
than the local reverberation sample, we expect that the size-luminosity relation will be tested at the higher luminosity regime, which is more relevant 
to measuring the \mbh\ of high-z QSOs.

\acknowledgments
We thank the anonymous referees for their comments, which improved the clarity of the manuscript. 
This work was supported by Samsung  Science and Technology Foundation under Project Number SSTF -BA1501-05
and the National Research Foundation of Korea grant funded by the Korea government (No. 2016R1A2B3011457).


\end{document}